\documentclass[letter,twocolumn]{jpsj2}

\newcommand{\simge}
{\raisebox{-0.75ex}[-1.5ex]{$\;\stackrel{>}{\sim}\;$}}
\def\1{1}

\def\e{{\epsilon}}
\def\k{{ {\bf k} }}

\def\q{{ {\bf q} }}

\def\w{{\omega}}
\def\a{{\alpha}}

\def\runauthor{Hiori {\sc Kino}, Hiroshi {\sc Kontani}, and Tsuyoshi {\sc Miyazaki}}

\title{
Phase Diagram of  $\beta'$-(BEDT-TTF)$_2$ICl$_2$ under High Pressure Based on the First-Principles Electronic Structure }
\author{
 Hiori {\sc Kino}$^1$, Hiroshi {\sc Kontani}$^2$, and Tsuyoshi {\sc Miyazaki}$^1$
}
\inst{
$^1$National Institute for Materials Science, 1-2-1 Sengen, Tsukuba, Ibaraki 305-0047, Japan.\\
$^2$Department of Physics, Saitama University, 255 Shimo-Okubo, Saitama-city, 338-8570, Japan.
}

\abst{
We present a theoretical study on the superconductivity of  $\beta'$-(BEDT-TTF)$_2$ICl$_2$ at $T_{\rm c}=$14.2~K under a high hydrostatic pressure recently found, which is the highest among organic superconductors. In the present work, we study an  effective model using the fluctuation-exchange (FLEX) approximation based on the results of first-principles calculation. In the obtained phase diagram, the superconductivity with $d_{xy}$-like symmetry is realized next to the antiferromagnetic phase, as a result of the one-dimensional to two-dimensional crossover driven by the pressure.
}

\begin{document}
\maketitle

  BEDT-TTF (ET) compounds can have many kinds of layered structures,
  each of which possesses a distinctive function.\cite{HF-approx} One of the most prominent functions is superconductivity, such as $\kappa$-(ET)$_2$X.\cite{Kanoda} It is shown that the temperature-pressure phase diagram of $\kappa$-(ET)$_2$X salt can be reproduced by the fluctuation-exchange (FLEX) approximation  \cite{Kino,Kondo,Schmalian}, which is a kind of self-consistent spin-fluctuation theory. \cite{Bickers,SCR} According to the FLEX approximation, $d$-wave superconductivity is expected and is mediated by the strong antiferromagnetic (AF) fluctuations due to the Coulomb interaction.

Recently, Taniguchi et al. discovered another superconductor, $\beta'$-(ET)$_2$ICl$_2$, whose superconducting (SC) transition temperature, $T_{\rm c}$, is 14.2~K under a high hydrostatic pressure ($P\simge$8.2~GPa). This is a new record $T_{\rm c}$ among organic superconductors. \cite{Taniguchi} At ambient pressure, $\beta'$-(ET)$_2$ICl$_2$ shows a semiconducting temperature dependence of conductivity below room temperature, and indicates a magnetic transition at $T_{\rm N}$=22~K. As pressure is applied, resistivity decreases gradually, and metallic behavior ($d\rho/dT >0$) is observed above $T_{\rm MIT}$ under 6.5~GPa. Note that $T_{\rm MIT}$, which is the crossover temperature, is expected to be higher than $T_{\rm N}$. At 8.2~GPa, the insulating phase (or $T_{\rm MIT}$) disappears and the SC transition occurs at $T_{\rm c}=14.2$~K at the same time.

At present, the atomic positions of $\beta$'-(ET)$_2$ICl$_2$ under applied pressure are unknown experimentally. In order to explain the electronic property in $\beta$'-(ET)$_2$ICl$_2$ under pressure, Kontani discussed the phase diagram of the title material in connection with the $\beta$-(ET)$_2$X, which is a superconductor ($T_{\rm c}=1\sim8$~K) with a two-dimensional Fermi surface  \cite{Kontani-beta}. He proposed that $d_{x^2-y^2}$-wave superconductivity due to AF fluctuations occurs, driven by the dimensional crossover under applied pressure.

After Kontani's study, the optimized atomic positions of $\beta$'-(ET)$_2$ICl$_2$ under applied pressure were calculated using the first-principles method based on the generalized gradient approximation. \cite{Kontani-beta,Miyazaki-ETICl2} At the same time, tight-binding parameters were fitted to reproduce the electronic structure near the Fermi level on the $b^*$-$c^*$ plane at several pressures. According to the result, the Fermi surface possesses two-dimensional nature at higher pressures as expected in ref. 9. However, its shape is different from that of  $\beta'$-(ET)$_2$ICl$_2$, contrary to the assumption in ref. 9.

In the present work, we study the origin of the superconductivity as well as the phase diagram in $\beta'$-(ET)$_2$ICl$_2$ based on the newly derived electronic structure from the first-principles calculation.\cite{Miyazaki-ETICl2} We employ the FLEX approximation based on the fact that the SC phase appears upon degradation of the AF phase by pressure. Although the obtained AF region is rather wider than that in experiments, the derived $T_{\rm c}$ is consistent with experiments on $\beta$'-(ET)$_2$ICl$_2$. The present analysis suggests that $d_{xy}$-wave superconductivity is realized due to spin fluctuations, ${\bf Q}=(q_y,q_z)\sim(\pi,0)$.

\begin{figure}
\includegraphics[width=8cm]{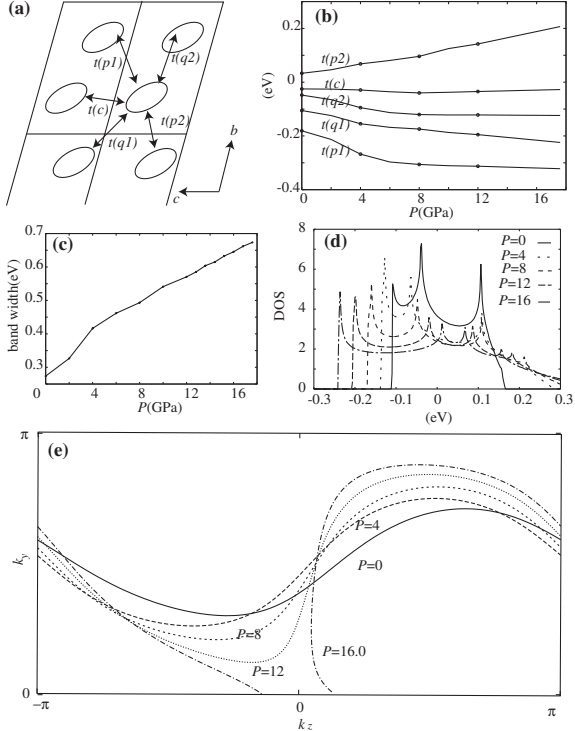}
\caption{
(a) Schematic crystal structure of $bc$-plane and transfer integrals between BEDT-TTF molecules. (b) Fitted transfer integrals (c) band width, (d) the density of states and (e) Fermi surfaces at applied pressures. The first-principles electronic structures are calculated structures at 0, 4, 8, and 12 Gpa. The transfer integrals at other pressures are interpolated using cubic spline or extrapolated linearly. The unit of $P$ (pressure) is GPa. 
}
\label{fig:model}
\end{figure}

Although the structures of organic materials are complex, it is well known that the band structure near the Fermi level is quite simple; it is sufficient to take into account the HOMOs of the ET molecules for the title material. The ET network is two-dimensional, reflecting the layered structure, and the schematic structures of $\beta'$-(ET)$_2$X are shown in Fig.~\ref{fig:model}(a). Each ellipse represents an ET molecule, and each highest occupied molecular orbital (HOMO) of the ET molecule possesses 1.5 electrons on average. The fitted transfer integrals are shown in Table~I and are plotted in Fig.~1(b).\cite{Miyazaki-ETICl2,comment-fit} As a good approximation, we take into account only the antibonding orbit of each pair of ET molecules connected by $t(p1)$, whose absolute value is much larger than the others. Then, the original system is well mapped onto the ``dimer model'' at half filling.  Hereafter, we study the dimer Hubbard model, whose relevant parameters are the tight binding integrals obtained by the fit to the upper HOMO band of the ET molecules, and the effective on-site Coulomb interaction ($U_{\rm eff}$) on the dimer molecule. 

First, let us review the electronic structure at $U=0$. In this study the pressure effect on the electronic states is only taken into account through the change in the transfer integrals: they are interpolated by the cubic spline up to a pressure of 12~GPa and extrapolated linearly over a pressure of 12~GPa. (See also Fig.~1(b).)  When one compares each transfer integral at ambient pressure with those at 4~GPa, the transfer integrals change greatly. However after 6~GPa, they hardly change except for the linear increase in  $t(p2)$, which is the counterpart of the intradimer transfer integral, $t(p1)$. As a result, the width of the upper band increases almost linearly after 4~GPa. (Fig.~1(c)) Such a feature is also confirmed in the density of states shown in Fig.~1(d).  The Fermi surfaces are also shown in Fig.~1(e) at a number of pressures. At an ambient pressure, the Fermi surface looks one-dimensional and the nesting is strong. However, at higher pressures, the Fermi surfaces become two-dimensional and a closed Fermi surface appears at $P$=16.0~GPa,  because of the upward shift of the saddle point near the $\Gamma$ point in the electronic structure. Note that the saddle point is located at the $\Gamma$ point, which results in the van Hove singularity in the density of states.  It is located just below the Fermi level at $P$=12.0~GPa and above the Fermi level at $P$=16.0~GPa. (also see  Fig ~1(d).) This peak gives a larger density of states than that expected from the bandwidth. This point will be discussed later.

\begin{table}
\caption{
 Fitted transfer integrals. Units of pressure and transfer integrals are GPa and eV, respectively.\cite{Miyazaki-ETICl2} See also Fig. 1(a) to specify the location of the transfer integrals in the crystal.}
\begin{center}
\begin{tabular}{cccccc} \hline 
$P$  & $ t(p1)$  &   $ t(p2)$   & $t(q1)$    &   $t(q2)$    &    $t(c )$ \\
\hline
 0 &  -0.181 &  0.0330 & -0.106 & -0.0481 & -0.0252 \\
4 &  -0.268 &  0.0681 & -0.155 & -0.0947 & -0.0291  \\
8 &  -0.306  & 0.0961 & -0.174 & -0.120 & -0.0399   \\
12 & -0.313 &  0.142 & -0.195 & -0.122 & -0.0347 \\ \hline
\end{tabular}
\end{center}
\end{table}

Next, in order to take into account the effects of electron correlations, $U_{\rm eff}$, we calculate the self-energy for the dimer model for $U_{\rm eff}$=0.25, 0.30 and 0.35~eV using the FLEX approximation  \cite{Bickers}. Hereafter, we write $U_{\rm eff}$ as $U$ for simplicity. To obtain the magnetic transition temperature $T_{\rm N}$, we calculate the Stoner factor without vertex corrections, $\a_{\rm S}$, given by 
\begin{eqnarray}
\a_{\rm S}= \max_{\k}\left\{ \ U\cdot \chi^{0}(\k,\w\!=\!0)\  \right\},
 \label{eqn:Stoner}
\end{eqnarray}
where   
\begin{eqnarray}
\chi^0(\q,\w_l)
 = -T\sum_{\k, n} G(\q+\k,\w_l+\e_n) G(\k,\e_n) \mbox{,}
     \label{eqn:chi0}
\end{eqnarray}
where $G(\q+\k,\w_l+\e_n)$ is the Green function given by the FLEX approximation, and $\w_l$ ($\e_n$) is the Matsubara frequency for boson (fermion). $T_{\rm N}$ is determined by the Stoner criterion, $\a_{\rm S}= \1$. In the FLEX approximation, however, $\a_{\rm S}$ does not exceed 1 at finite temperatures in two-dimensional systems, which is consistent with the Mermin and Wagner theorem. So we determine $T_{\rm N}$ under the condition $\a_{\rm S}= \a_{\rm N}$, where we set $\a_{\rm N}$ as $(1-\a_{\rm N})^{-1} \sim O(100)$. The AF state will occur through the weak coupling between layers,  $J_\perp$.  \cite{AF-condition}  
To evaluate the SC transition temperature $T_{\rm c}$, We solve the linearized Eliashberg equation. For a singlet-pairing case [$\phi(-\k,\e_n)= + \phi(\k,\e_n)$], it is given by \cite{2D-SC-Monthoux} \begin{eqnarray} \lambda \cdot \phi(\k,\e_n) &=& -T\sum_{\q, m}  V(\k-\q,\e_n-\e_m)  \nonumber \\ & & \times G(\q,\e_m) G(-\q,-\e_m)  \cdot \phi(\q,\e_m), \label{eqn:lambda} \end{eqnarray} where ${V}(\k,\w_l)= \frac32 U^2 \frac{{\chi}^0(\k,\w_l)}{1-U{\chi}^{0}(\k,\w_l)} - \frac12 U^2 \frac{{\chi}^0(\k,\w_l)}{1+U{\chi}^{0}(\k,\w_l)}   + U$. $T_{\rm c}$ is given by the condition $\lambda=1$. In the FLEX approximation, a finite $T_{\rm c}$ is obtained even in two-dimensional systems irrespective of the Hohenberg theorem. However, this approximation gives reasonable $T_{\rm c}$'s for $\kappa$-(ET) organic compounds and high-$T_{\rm c}$ cuprates  \cite{Kino,Kondo,Schmalian,2D-SC-Monthoux}. In these calculations, we adopted 64$\times$64 $k$-points and 512 Matsubara frequencies, and executed summations on frequency and wave vectors by employing fast Fourier transform techniques.

\begin{figure}
\includegraphics[width=8cm]{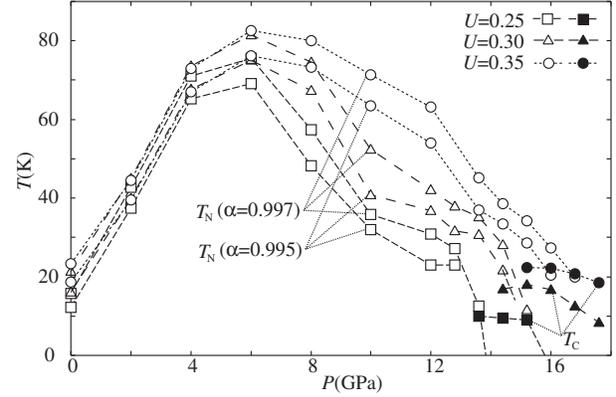}
\caption{N\'eel temperature ($T_{\rm N}$) and superconducting transition temperature ($T_{\rm c}$) as functions of pressure. $T_{\rm N}$ is evaluated by the Stoner criterion, $\a_{\rm S}$=0.995 or 0.997. Lines are guides for the eyes. The unit of $U$ is eV. }
\label{fig:Phase}
\end{figure}
 
The obtained phase diagram is given in Fig.~\ref{fig:Phase}. First let us see the N\'eel temperature. At $U=$0.30~eV, the $T_{\rm N}$ at ambient pressure is 16~K. With increasing pressure, $T_{\rm N}$ increases rapidly and has a peak at 75~K at 6~GPa. With further increase in pressure, $T_{\rm N}$ decreases gradually and finally $T_{\rm N}$ becomes smaller than $T_{\rm c}$ at 15.2~GPa and the antiferromagnetism is substituted for the superconductivity in the phase diagram. A similar trend is also observed at other $U$s; e.g., $T_{\rm N}$ is 12~K at ambient pressure at U=0.25~eV, has a maximum value at 69~K at 6~GPa and decreases under further applied pressure and is finally replaced with $T_{\rm c}$. A larger $U$ yields a larger $T_{\rm N}$ as well as a wider area of the antiferromagnetic phase. 

Next we discuss the obtained magnetic order. The nesting vector $Q$ where $\chi(Q,\w)$ has a maximum is commensurate and $(\pi, 0)$ for the low-pressure region, $P\le$12~GPa for $U$=0.35~eV ($P\le$8~GPa for $U$=0.30 and $U$=0.25~eV). This is a result  of the strong electron correlation given by the FLEX approximation; in fact, $Q$ takes an incommensurate value at $U=0$. For higher-pressure regions, on the other hand, $Q$ becomes incommensurate and is $~(\pi-0.125\pi, 0)$ when $T_{\rm c}$ has a maximum. A shallow dip in $T_{\rm N}$ at approximately 14~GPa for $U$=0.35 (10~GPa for $U$=0.25 and $U$=0.30) corresponds to the crossover pressure between the commensurate and incommensurate $Q$.

Finally let us discuss the superconductivity. If the Stoner criterion is 0.995,
 the maximum $T_{\rm c}$ under the condition $T_{\rm c}>T_{\rm N}$ for $U=0.30$~eV is approximately 18~K at 15.2~GPa. It decreases as pressure increases, and $T_{\rm c}\sim 8$~K at 17.6~GPa.  The same feature can be seen in the case of $U$=0.25~eV and 0.35~eV except for the shift in  the transition temperatures.  A smaller $U$ gives a smaller $T_{\rm c}$ as well as a smaller transition pressure between antiferromagnetism and superconductivity. The maximum $T_{\rm c}$ is 9.5~K(18.5~K) at 14.4~GPa(17.6~GPa) for $U$=0.25~eV(0.35~eV). 

\begin{figure}
\includegraphics[width=8cm]{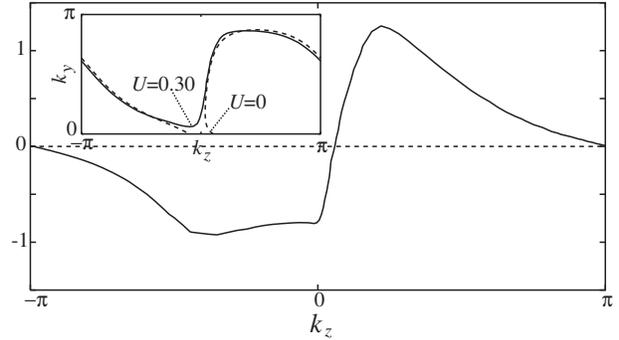}
\caption{
$\phi(k_y,k_z,i\pi T)$ on the Fermi surface at U=0.30~eV, T=14.4~K and the pressure of 15.2~GPa. An inset shows Fermi surfaces of U=0 (a dotted line) and U=0.30~eV (a solid line) at 15.2~GPa. The unit of $U$ is eV. }
\label{fig:Cph}
\end{figure}

Figure \ref{fig:Cph} shows the solution of eq. (\ref{eqn:lambda}) at $(k_y,k_z)$ on the Fermi surface. The inset shows the change of the Fermi surface at $U$=0 and $U$=0.30~eV.  The saddle point near the $\Gamma$ point causes a peak in the DOS above the Fermi level at $U$=0, however, it is below the Fermi level at $U$=0.3~eV because of the $k$-dependence of the self-energy ($\Sigma_k(0)$). As a result, the Fermi surface for $U$=0.3~eV is still open at this pressure. 

The main figure shows that the obtained SC order parameter has line nodes (along $a$ direction) as in the high-$T_{\rm c}$ cuprates and the $\kappa$-(ET) compounds. However, the nodes exist at $k_z=\pm \pi$ and $k_z\sim 0$, therefore the SC order parameter is $d_{xy}$-wave-like judging from the position of the nodes. Note that the $b$-$c$ coordinate in the figure is square and is different from that of the original crystal for simplicity.

Here we discuss the physical meanings of the calculated results. The obtained $T_{\rm N}$ increases with increasing pressure up to 6~GPa. This behavior is totally different from the result obtained by the mean field approximation, which predicts a high  $T_{\rm N}$ when the nesting condition is good and when DOS at the Fermi surface for $U=0$ (i.e., the inverse of the bandwidth) is large. This discrepancy comes from the fact that DOS at the Fermi level for finite $U$ is reduced due to the singular $\omega$ and $k$-dependence of the self-energy  $\Sigma_k(\omega)$ given by the FLEX approximation. The reduction in DOS is prominent as the Fermi surface becomes one-dimensional-like, because the singularity of $\Sigma_\k(\omega)$ due to spin fluctuations becomes prominent in lower-dimensional systems \cite{Kino-TMTSF}. In other words, the increase in $T_{\rm N}$ is a natural consequence of the dimensional increase as stressed in the previous paper.\cite{Kontani-beta} As a result, the fine nesting does not always give higher $T_{\rm N}$ contrary to the present of the spin density wave in weakly correlated systems because strong fluctuations in the lower-dimensional systems, or at lower pressures in this case,  suppress growth on the long-range order. This is natural and plausible, and one of the merits of the FLEX approximation. 

Another important effect of the self-energy is that the nesting vector tends to be pinned to the commensurate value. The nesting vector at $U=0$ is not $(\pi, 0)$, however, it is $(\pi, 0)$ for moderate $U$ owing to the k-dependence of $\Sigma_k(0)$ by the FLEX approximation. This effect naturally connects the magnetic structure in the strongly correlated (Mott) insulators with that in the strongly correlated metals. The former is always commensurate, on the other hand, the latter can be incommensurate, but it becomes commensurate because of the strong  electron correlations.  It is also connected with the distortion of the Fermi surface at finite U; the electron correlations tend to make the Fermi surface nest better. 

 With further increase in pressure, the nesting condition deteriorates and the nesting vector becomes incommensurate. Then, $T_{\rm N}$ decreases and finally the antiferromagnetic phase gets totally suppressed with increasing pressure. The SC phase that appeared in the high-pressure region is next to the antiferromagnetic phase and its SC order parameter has $d$-symmetry. The origin of the superconductivity is considered to be the AF fluctuation similar to those of (TMTSF)$_2$X and $\kappa$-(ET)$_2$X. As a result, the obtained phase diagram, Fig.~\ref{fig:Phase}, as well as the concept of the superconductivity driven by the dimensional crossover, will make sense and be reasonable.

When one compares Fig.~2 with the experimental phase diagram, one notices that the pressure where the antiferromagnetism disappears is much larger than the experimental one. Here we enumerate the possible origins of this discrepancy: the error of the first-principles calculation, the approximation to calculate the self-energy and the model itself. 

    The accuracy of the first-principles calculations for the title compound is investigated in ref.~10.  Although it is well known that first-principles calculations based on the generalized gradient approximation cannot reproduce the van der Waals interactions correctly, the investigation shows that the error of calculated pressure is approximately 0.6 Gpa, which will not change the results shown in the present work. Similar results are reported for molecular solids and other charge-transfer molecular solids. \cite{Tateyama-pressureLDA,Miyazaki-dmit}

Though it should be experimentally confirmed, we expect that the first-principles calculation can determine the crystal and electronic structures of the title compound fairly accurately. Thus, one of the origins of the discrepancy on the transition pressure might be traced to the FLEX approximation. In a similar calculation of a nearly one-dimensional case at half-filling such as TMTSF salts, the SC phase is almost concealed with the AF phase in the FLEX approximation. \cite{Kino-TMTSF} The particle-particle scattering channel or the vertex correction for susceptibility, which is not taken into account in the present approximation, may decrease the excessive AF region. 

We also comment on several problems concerning the effective model. We fixed $U$ in the obtained phase diagram for simplicity, however, such an assumption is violated in real systems because the value of $t(p1)$ considerably increases between $P$=0 and $P$=4~GPa.  Although $t(p1)$  is approximately constant for higher pressures, another problem  emerges. The absolute value of $t(p2)$ is considerably enhanced and becomes compatible with $|t(p1)|$ under high pressures, which means that  the assumption of the dimer model, $|t(p1)|\gg |t(p2)|$, deteriorates at a higher-pressure region. Therefore,  $U_{\rm eff}$ at higher pressures is probably overestimated. As a result, $U_{\rm eff}$, which is reasonably estimated to be $~0.36$~eV at ambient pressure from the value of $t(p1)$,  increases with pressure below 4~GPa.   $U_{\rm eff}~0.6$~eV at 4~GPa, and it decreases for much higher pressures.  This effect will make $T_{\rm N}$ and $T_{\rm c}$ smaller and reduce the critical pressure for the superconductivity. In addition, the dimer Hubbard model overestimates the spin fluctuation somehow, which will stabilize the AF phase too much. By solving these problems, we could obtain a theoretical phase diagram closer to the experimental one.

Finally let us consider the reason why this system has a high $T_{\rm c}$. The bandwidth increases almost linearly with increasing pressure as shown in Fig.~1. However DOS at the Fermi level does not decrease reciprocally as a function of pressure. There exists a saddle point near the $\Gamma$ point that causes the peak (van-Hove singularity) in DOS. This peak shifts upward with increasing pressure, and the tail of the peak enlarges DOS near the chemical potential. As a result, DOS at the Fermi level at $U$=0 is almost constant, thus this system has a larger DOS at a chemical potential than the one expected from the bandwidth for $P\ge$4~GPa, which includes the pressure of experimental SC and theoretical SC. This finding is related to the high experimental $T_{\rm c}$. We note that $T_{\rm c}$ in a quasi-one-dimensional system free from the van-Hove singularity near the Fermi level is very low ($\approx1$~K), which is recognized by the FLEX approximation  \cite{Kino-TMTSF}. 

In summary, we studied the origin of the superconductivity in $\beta'$-(ET)$_2$ICl$_2$ based on the FLEX approximation, using the tight binding fit to the electronic structure in the first-principles calculation at ambient and under applied pressures. According to the present analysis, a wide AF insulating phase is realized in the lower pressures region. Its magnetic ordering vector is approximately $(\pi,0)$.  For the higher pressure region, next to the AF insulating phase, the  $d_{xy}$-wave superconductivity occurs owing to the AF fluctuations approximately $(\pi,0)$.  High $T_{\rm c}$ in $\beta'$-(ET)$_2$ICl$_2$ is realized owing to the one-dimensional to two-dimensional crossover as a result of the  crystal structure change under high pressures.  Experimental checks for the magnetic ordering vector as well as the symmetry of the SC order parameter are highly demanded.

The authors are grateful to H. Taniguchi and Y. Tateyama for valuable discussions. This work is partly supported by Special Coordination Funds of MEXT of the Japanese Government, and ACT-JST.


\begin{thebibliography}{99}
\bibitem{comment-fit} All the electronic structures are fitted fairly well at ambient pressure. At higher pressures, the fitting becomes worse and the electronic structures cannot be fitted to reproduce all the dispersion relations, therefore, they are fitted to reproduce the dispersions and Fermi surfaces especially near the Fermi levels. The bottom of the upper HOMO band is not fitted well especially at higher pressures, because the electronic structure of the top of the lower HOMO band mixes with that of the bottom the upper HOMO band.
\bibitem{AF-condition} 
 The AF order will occur under the condition that
 $1-J_\perp \chi(Q,0)=0$. That is, $J_\perp = U(1-\a_{\rm S})\sim O$(10 K),
 which is realistic for real materials.
\end{thebibliography}
\end{document}